\documentclass[9pt,twocolumn,twoside]{osajnl}

\journal{ol} 

\setboolean{shortarticle}{true}

\title{Applying tiling and pattern theory in the design of hollow-core photonic crystal fibers for multi-wavelength beam guidance}

\author[1,*]{Z. Montz}
\author[1]{A. A. Ishaaya}

\affil[1]{School of Electrical and Computer Engineering, Ben-Gurion University of the Negev, Beer Sheva, 8410501, Israel}

\affil[*]{Corresponding author: zevm@post.bgu.ac.il}

\ociscodes{(060.5295) Photonic crystal fibers; (060.2280) Fiber design and fabrication; (190.4370) Nonlinear optics, fibers; (190.2620) Harmonic generation and mixing.}

\usepackage{amsmath,amssymb,graphicx,physics,caption,gensymb}
\usepackage{textcomp} 
\usepackage{cleveref} 
\usepackage{siunitx} 
\usepackage{xfrac} 

\crefformat{equation}{(#2#1#3)} 

\hypersetup{
  colorlinks   = true, 
  urlcolor     = blue, 
  linkcolor    = blue, 
  citecolor    = blue  
}

\begin{abstract}
We show how to apply tiling and pattern theory in the design of hollow-core photonic crystal fibers and guide light in multiple bandgaps. By combining two different glass apexes in a single~[3\textsuperscript{6};3\textsuperscript{2}.4.3.4] 2-uniform tiling, transmission regions with fundamental, second and third harmonic wavelengths are supported. This cladding design may also be an excellent candidate for high power beam delivery of Er/Yb, Nd:YAG and Ti:Sapphire laser sources.
\end{abstract}

\setboolean{displaycopyright}{true}

\begin{document}

\setlength{\parskip}{0pt} 
\setlength{\textfloatsep}{6pt}

\setlength\abovedisplayskip{-5pt}
\setlength\belowdisplayskip{5pt}
\setlength\abovedisplayshortskip{-5pt}
\setlength\belowdisplayshortskip{5pt}

\maketitle

\noindent
For the past two decades, hollow-core photonic crystal fibers (HC-PCFs)~\cite{birks1995full,cregan1999single} have had a tremendous impact on nonlinear optics. HC-PCFs have allowed to reduce the threshold of many nonlinear processes by several orders of magnitude~\cite{benabid2002stimulated}. This dramatic threshold reduction was obtained by tight confinement of the fundamental mode (FM), and by increasing dramatically the light-gas interaction length. HC-PCFs that guide a FM within a certain range of frequencies and reject frequencies outside this range are termed hollow-core photonic bandgap fibers (HC-PBGFs). The HC-PBGFs FM confinement loss is less affected by material absorption since the fiber core is hollow and does not contain a glass matrix composite. It was shown~\cite{roberts2005ultimate} that the FM confinement loss within the bandgap is limited by scattering losses at the core boundary.

Many nonlinear processes, such as high harmonic generation, require multiple bandgaps that will guide the laser pump signal and generated harmonics with low confinement loss. It was shown that PBGFs with interstitial holes~\cite{broeng1998highly} and high air-filling structures~\cite{mortensen2004modeling,yan2005design,light2009double,lyngso20097} can guide light in two separate bandgaps. Yet, the ratio between their central normalized frequencies at the air line~$ck_z/w=1$ is not suitable for second harmonic~(SH) and third harmonic~(TH) guidance. Recently it was theoretically demonstrated that HC-PBGFs can guide light in two well separated bandgaps suitable for TH~\cite{montz2015dual} guidance. Such cladding designs can guide the FM and TH Gaussian modes with low confinement loss; yet, it is unclear if phase-matching of these modes is feasible.

Several approaches have been proposed to phase-match second harmonic generation~(SHG) and third harmonic generation~(THG) in PCFs. In solid core PCFs, phase-matching a Gaussian total internal reflection mode with a Gaussian bandgap mode was theoretically proposed~\cite{betourne2008design} and experimentally demonstrated~\cite{cavanna2016hybrid}. In HC-PCFs, it was theoretically proposed to quasi phase-match high harmonic generation by modulating the phase of ionization electrons using a counter-propagating beam~\cite{cohen2007grating,ren2008quasi}. THG was experimentally demonstrated in an Ar-filled kagome fiber by counterbalancing the fiber dispersion and the gas dispersion with two different order modes~\cite{nold2010pressure}. SHG was experimentally demonstrated in a Xe-filled kagome fiber by applying an external DC field~\cite{menard2015phase}. Kagome fibers~\cite{benabid2002stimulated} are an excellent platform for high harmonic generation since their transmission region is broadband.

Here we show for the first time, to the best of our knowledge, how to apply tiling and pattern theory~\cite{grnbaum1987tilings} in the cladding of high air-filling HC-PCFs and guide light in multiple bandgaps. With this approach it is possible to design cladding structures that can guide a FM, SH and TH simultaneously. All of the modes, which are guided in separate bandgaps, have Gaussian distributions; therefore, their spatial overlap is exceptionally good.

\begin{figure}[!t]
\centering
\includegraphics[width=0.63\linewidth]{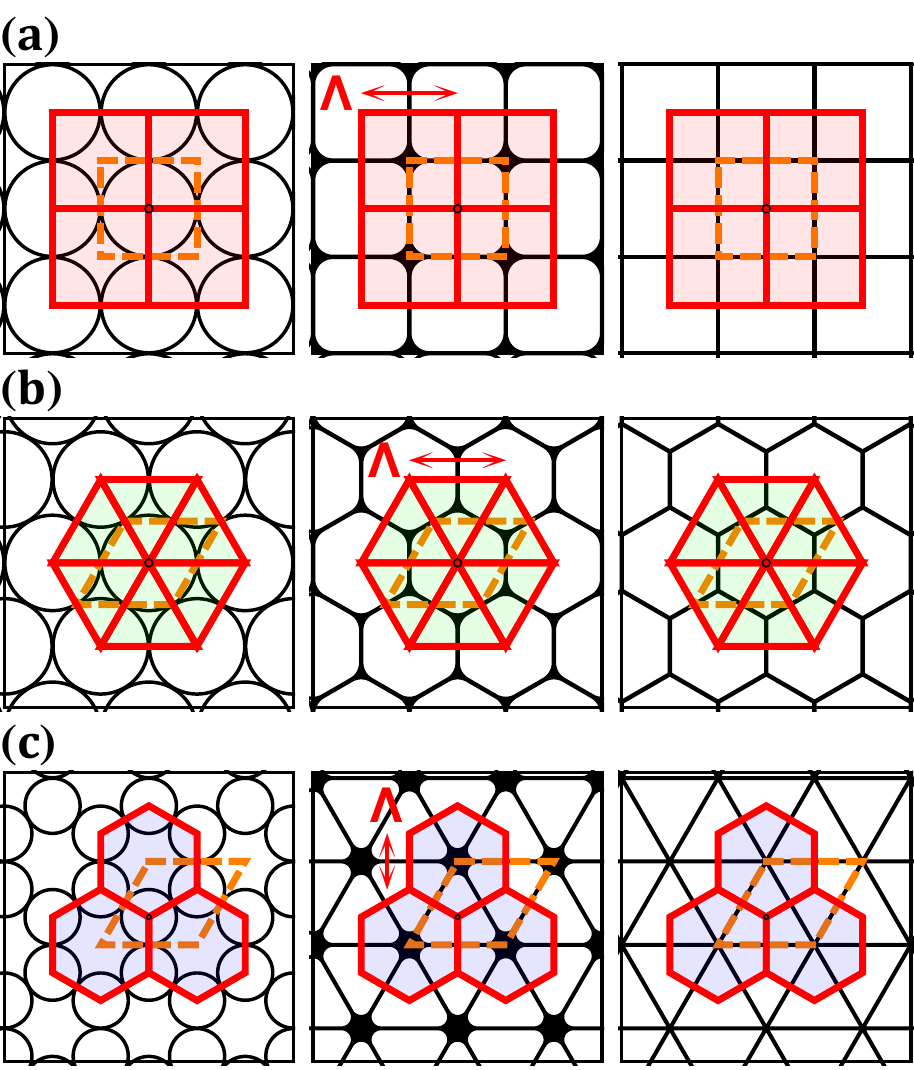}
\captionsetup{justification=justified}
\caption{(a) Square, (b) triangular and (c) hexagonal uniform tilings. In all columns, the tilings (polygons with red edges) surrounding each vertex and the unit cells (polygons with orange dashed edges) of the structures are shown. Left column shows the preforms of the high air-filling structures, center column shown the high air-filling structures after pressurizing the thin capillaries during the fiber draw and right column shows the Laves tilings. Apexes are located at the center of the tilings and struts are perpendicular to the tilings' edges.}
\label{fig:1}
\end{figure}

Pressurizing preform thin glass capillaries, which are centered on the vertexes of uniform tilings~(Fig.~\hyperref[fig:1]{\ref*{fig:1}}, left column), creates high air-filling structures with apexes that are located at the center of the tilings and glass struts that are perpendicular to the tilings' edges (Fig.~\hyperref[fig:1]{\ref*{fig:1}}, center column). When the apex curvatures of high air-filling structures are equal to zero, we simply obtain the Laves tilings~(Fig.~\hyperref[fig:1]{\ref*{fig:1}}, right column). Throughout the manuscript, apex curvatures are identical to the definitions in~\cite{poletti2010hollow} and structures are termed according to their tilings. Apex separation in the triangular, square and hexagonal structures are~${\Lambda/\sqrt{3}}$, ${\Lambda}$ and~${\sqrt{3}\Lambda}$, respectively. ~${\Lambda}$ is the distance between the center of adjacent capillaries, which is equal to the tilings' edges, and also equal to the square and triangular unit cell edge length~(Figs.~\hyperref[fig:1]{\ref*{fig:1}(a)} and~\hyperref[fig:1]{\ref*{fig:1}(b)}, center column).

\begin{figure}[!t]
\centering
\includegraphics[width=0.63\linewidth]{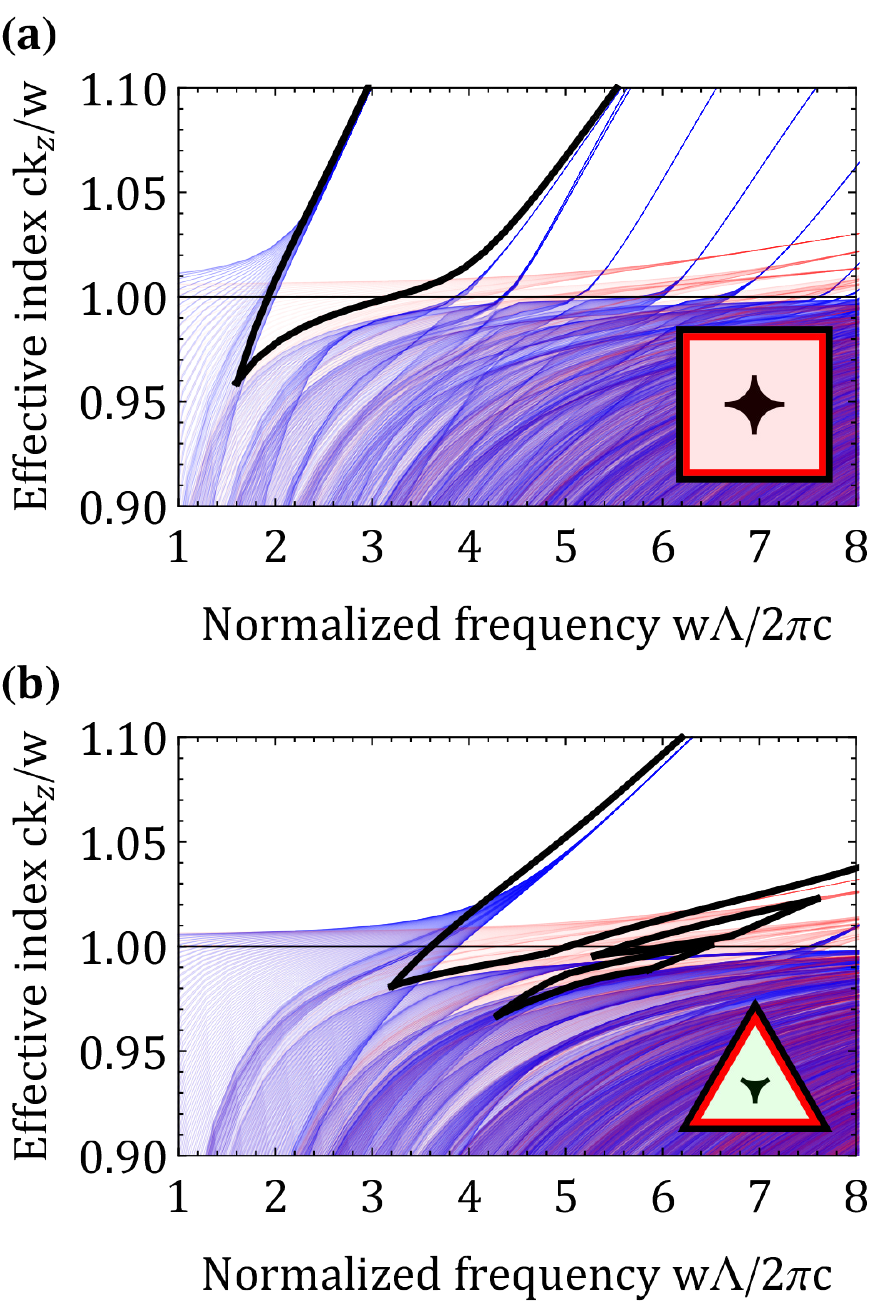}
\captionsetup{justification=justified}
\caption{Separate band diagrams of the apexes (blue) and struts (red) in the (a) square and (b) triangular structures with an~${0.15\Lambda}$ apex curvature and~${0.01\Lambda}$ strut thickness. Bandgaps of the structures are shown in bold (black), horizontal line (black) is the air line. Figure inset shows the structures' apexes, which are located at the center of the tilings.}
\label{fig:2}
\end{figure}

Fig.~\hyperref[fig:2]{\ref*{fig:2}} shows the struts and apexes band diagrams of the square and triangular structures. Band diagrams were computed with the plane wave expansion method~(PWE)~\cite{johnson2001block,guo2003simple}. Glass dielectric constant was set to~${2.1}$, apex radius curvature and strut thickness were set to~${0.15\Lambda}$ and~${0.01\Lambda}$, respectively. Fixing the dielectric constant is useful since it allows to find the band diagrams in a scalable system and scale the structure according to the desired nonlinear process of interest~\cite{joannopoulos2008photonic}. Fused silica low dispersion allows to preserve quite well the ratio between the central frequencies of the bandgaps during the scaling process. The structures' fundamental bandgap lowest normalized frequency are located near the apexes' fundamental bandgap lowest normalized frequency. The struts' band diagrams span the entire apexes' fundamental bandgap and are known to reduce the frequency bandwidth of the structures' bandgaps~\cite{light2009double}. In the square structure, the struts slightly narrow the apex fundamental bandgap~(Fig.~\hyperref[fig:2]{\ref*{fig:2}(a)}); yet, in the triangular structure~(Fig.~\hyperref[fig:2]{\ref*{fig:2}(b)}) the struts causes the apex fundamental bandgap to be discontinuous with several bandgaps. In the triangular structure, only the first higher order bandgap has a substantial frequency bandwidth at the air line. Properties of the first higher order bandgap were investigated rigorously in~\cite{light2009double} with different apex radius curvatures and strut thicknesses. Total bandwidth of the apexes' fundamental bandgap is reduced dramatically in the triangular structure compared with the square structure. Fig.~\hyperref[fig:2]{\ref*{fig:2}} demonstrates that while the apex band diagrams can approximate the lowest normalized frequency of the structures' fundamental bandgap, they cannot predict accurately the frequency bandwidth of the structures' bandgaps.

\begin{figure}[!t]
\centering
\includegraphics[width=0.63\linewidth]{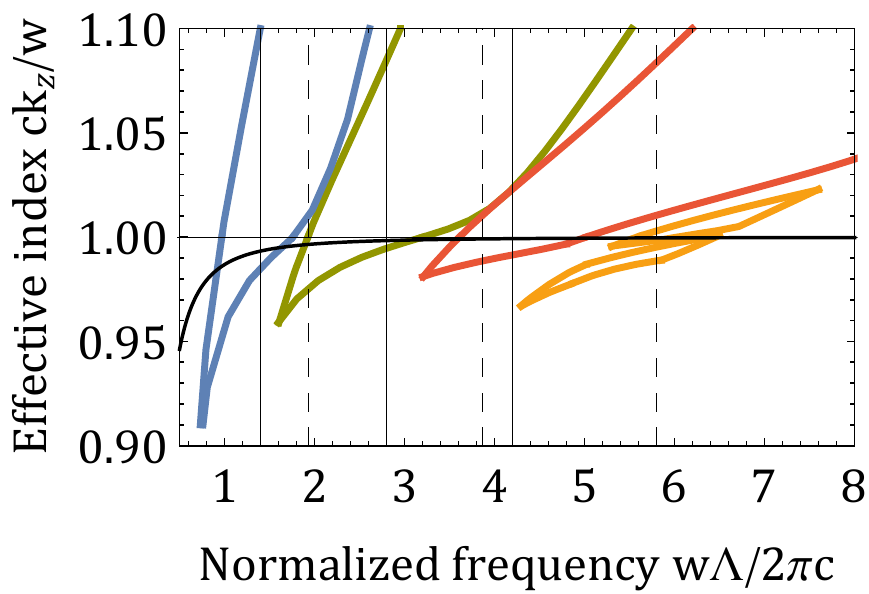}
\captionsetup{justification=justified}
\caption{Fundamental bandgaps of the hexagonal (blue), square (green) and triangular (red) structures with an~${0.15\Lambda}$ apex curvature and~${0.01\Lambda}$ strut thickness. The triangular structure has two higher order bandgaps (yellow); yet, only the first higher order bandgap has a substantial frequency bandwidth. The SH and TH could be supported at normalized frequencies of~${4.2}$, ${2.8}$ and~${1.4}$ by combining the properties of the hexagonal, square and triangular structures (black vertical lines). The SH and TH could also be supported at normalized frequencies of~${5.8}$, ${\sim3.87}$ and~${\sim1.93}$ by combining the properties of the square and triangular structures (black vertical dashed lines). For reference, the FM effective index of a capillary~\cite{marcatili1964hollow} is shown for wavelength and core diameter of~${1500~nm}$ and~${14.2~\mu m}$, respectively.}
\label{fig:3}
\end{figure}

Band diagrams of the hexagonal, square and triangular structures with an~${0.15\Lambda}$ apex curvature and~${0.01\Lambda}$ strut thickness are shown in~Fig.~\hyperref[fig:3]{\ref*{fig:3}}. High air-filling structures with a smaller apex separation have a fundamental bandgap at a higher normalized frequency. If somehow several properties of these high air-filling structures could be combined there would be two obvious approaches to guide the SH and TH. The first approach would be to combine the guiding properties of the hexagonal, square and triangular structures. The hexagonal fundamental bandgap would guide the FM, the square and triangular fundamental bandgaps would guide the SH and TH. This approach has two disadvantages: (a) supporting the FM and the first two harmonics requires a structure with three different apexes. Since each apex is expected to reduce the frequency bandwidth of the other apex bandgaps as a result of interference effects, reducing the number of different apexes in the structure is preferred for high harmonic guidance. (b) the FM will be guided at a low normalized frequency, thus will have a high confinement loss in the fundamental bandgap. This high confinement loss could be reduced by increasing the number of periods in the cladding~\cite{montz2015dual} or by designing a customized core~\cite{poletti2007hollow}; yet, these methods will make the fabrication of the fiber much more difficult. The second approach would be to combine the guiding properties of the square and triangular properties and guide the FM in the square fundamental bandgap and use the triangular fundamental and first higher order bandgaps to guide the SH and TH. This approach eliminates the disadvantages mentioned previously: (a) the structure will have only two different apexes instead of three, which will reduce apex interference effects and make the fabrication of the fiber easier and (b) the FM will be guided in the fundamental square bandgap at a higher normalized frequency, thus reducing the confinement loss of the FM.

Since the fundamental bandgap is shifted to lower normalized frequencies as the apex separation increases, and since guiding with a low confinement loss is difficult when the fundamental bandgap is at a low normalized frequency, it is preferred to construct hybrid structures that only have hexagonal, square and triangular tilings and avoid more complicated tilings with larger apex separation such as dodecagons. It is also preferred to use tilings that are invariant for~${60\degree}$ rotations (such as the p6m symmetry tilings) since they will allow to construct a circular-shaped core much more easily. The square and triangular apexes are located at the center of the square and triangular tilings and therefore, obtaining the guiding properties of the square and triangular structure could be realized in a single structure by simply placing the center of the preform capillaries on the vertices of a hybrid structure with square and triangular tilings. The higher order bandgap of the triangular structure is generated with struts of length~${\Lambda/\sqrt{3}}$, thus enforcing another limitation on the desired hybrid structure.

\begin{figure}[!t]
\centering
\includegraphics[width=0.63\linewidth]{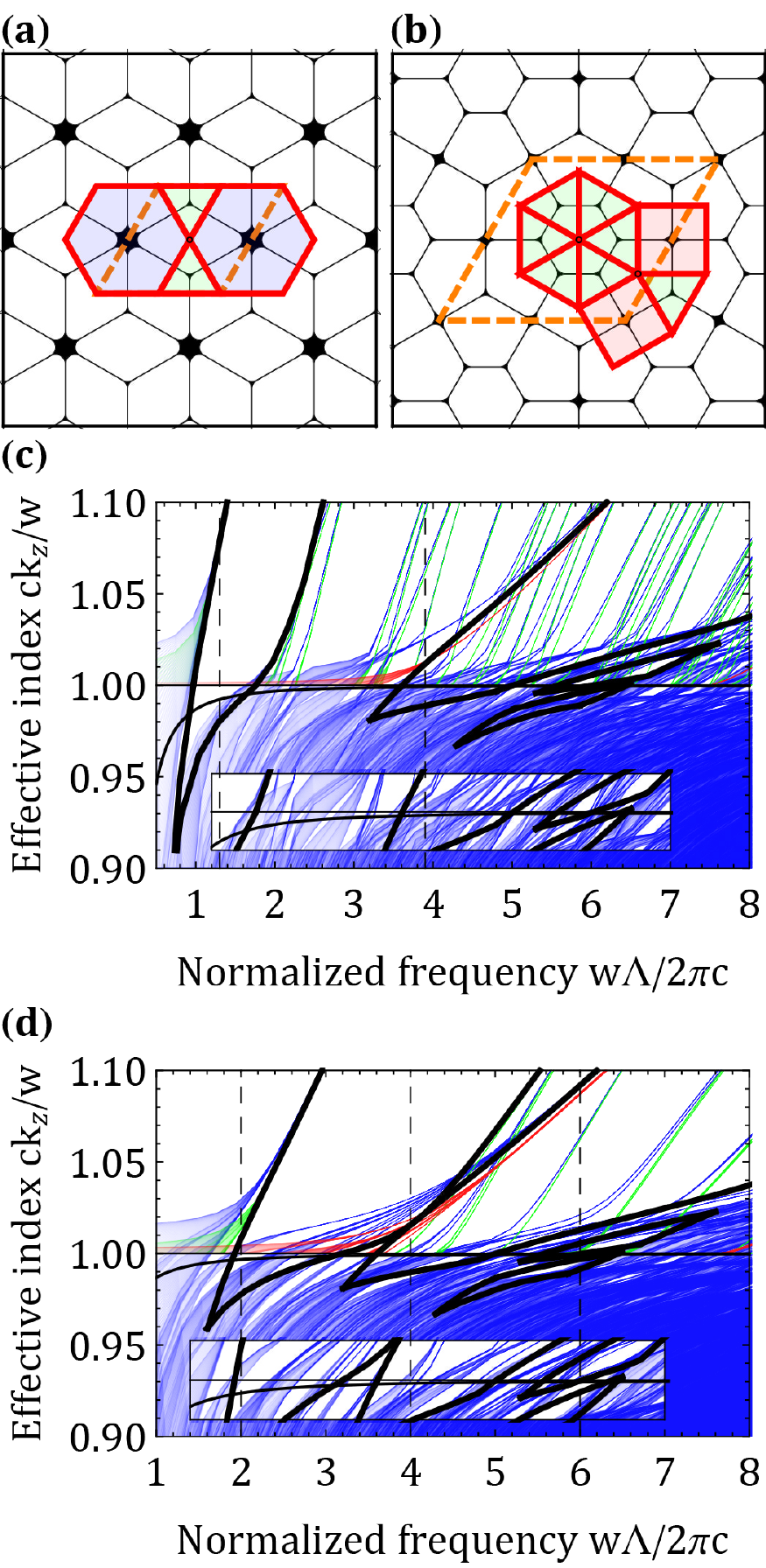}
\captionsetup{justification=justified}
\caption{(a) Trihexagonal uniform tiling and (b) [3\textsuperscript{6};3\textsuperscript{2}.4.3.4] 2-uniform tiling, and their corresponding high air-filling structures. Band diagrams of the (c) trihexagonal and (d) [3\textsuperscript{6};3\textsuperscript{2}.4.3.4] structures. In (c) green and red curves are the band diagrams of the hexagonal and triangular apexes, respectively. In (d) green and red curves are the band diagrams of the square and triangular apexes, respectively. Bandgaps of the hexagonal, square and triangular structures (see~Fig.~\hyperref[fig:3]{\ref*{fig:3}}) are shown in black bold. Figure inset shows the structures' band diagrams within the effective index range of $1 \pm 0.01$.}
\label{fig:4}
\end{figure}

Out of all eleven uniform tilings, none have only square and triangular tilings with a p6m symmetry; yet, there may be a uniform tiling that supports the TH. The trihexagonal tiling~(Fig.~\hyperref[fig:4]{\ref*{fig:4}(a)}) is a good candidate since the hexagonal fundamental bandgap can guide the FM and the triangular fundamental bandgap can guide the TH. The trihexagonal structure has struts of length~${2\Lambda/\sqrt{3}}$ and does not have struts of length~${\Lambda/\sqrt{3}}$; yet, the triangular higher order bandgap is not necessary for guiding the TH with this structure. Out of all 2-uniforn tilings~\cite{grnbaum1987tilings,krotenheerdt1968homogenen} the~[3\textsuperscript{6};3\textsuperscript{2}.4.3.4] tiling~(Fig.~\hyperref[fig:4]{\ref*{fig:4}(b)}) fulfills all restrictions mentioned above. The struts in the structure have two different lengths: ${\Lambda/\sqrt{3}}$, and~${\Lambda(1+\sqrt{3})/2\sqrt{3}}$. Struts of length~${\Lambda/\sqrt{3}}$ should support the higher order bandgap of the triangular structure; yet, it is unclear if the struts of length~${\Lambda(1+\sqrt{3})/2\sqrt{3}}$ will prevent the utilization of this higher order bandgap as a result of interference effects. Band diagrams of the trihexagonal and~[3\textsuperscript{6};3\textsuperscript{2}.4.3.4] structures with~${0.15\Lambda}$ apex curvature and~${0.01\Lambda}$ strut thickness are shown in~Figs.~\hyperref[fig:4]{\ref*{fig:4}(c)} and~\hyperref[fig:4]{\ref*{fig:4}(d)}.

Several interesting properties are shown in the band diagram of the trihexagonal structure. First, the hexagonal apexes' fundamental bandgap is not substantially affected by the triangular apexes in the structure; yet, the triangular apexes' fundamental bandgap is almost completely blocked by the hexagonal apex modes interference effects. This statement is validated by separating the hexagonal and triangular apexes' band diagrams in the trihexagonal structure (Figs.~\hyperref[fig:4]{\ref*{fig:4}(c)}, red and green curves). Many hexagonal apex modes are crossing through the triangular apex fundamental bandgap. Second, the hexagonal and triangular apexes' fundamental bandgaps lowest normalized frequencies are very similar to the triangular and hexagonal fundamental bandgaps lowest normalized frequencies (Figs.~\hyperref[fig:4]{\ref*{fig:4}(c)}, bold curves). Third, the triangular higher order bandgap is also blocked. This is expected since the trihexagonal structure does not have struts of length~${\Lambda/\sqrt{3}}$. Overall, the trihexagonal structure does not support the TH with~${0.15\Lambda}$ apex curvature and~${0.01\Lambda}$ strut thickness.

The~[3\textsuperscript{6};3\textsuperscript{2}.4.3.4] structure band diagram has much more promising results for high harmonic guidance. Although there are several triangular apex modes crossing through the square fundamental bandgap, there is still a bandgap near the normalized frequency of~${\sim2}$. Similarly, several square apex modes are crossing through the triangular fundamental bandgap; yet they do not completely block it, and there are still several small bandgaps near the SH. An important property of the~[3\textsuperscript{6};3\textsuperscript{2}.4.3.4] structure is that it has struts of length~${\Lambda/\sqrt{3}}$ and therefore, supports the TH near the triangular first higher order bandgap. The struts of length~${\Lambda(1+\sqrt{3})/2\sqrt{3}}$ do not block this bandgap with interference effects. Thus, unlike the trihexagonal structure, the~[3\textsuperscript{6};3\textsuperscript{2}.4.3.4] structure can support both the SH and TH.

\begin{figure}[!t]
\centering
\includegraphics[width=0.63\linewidth]{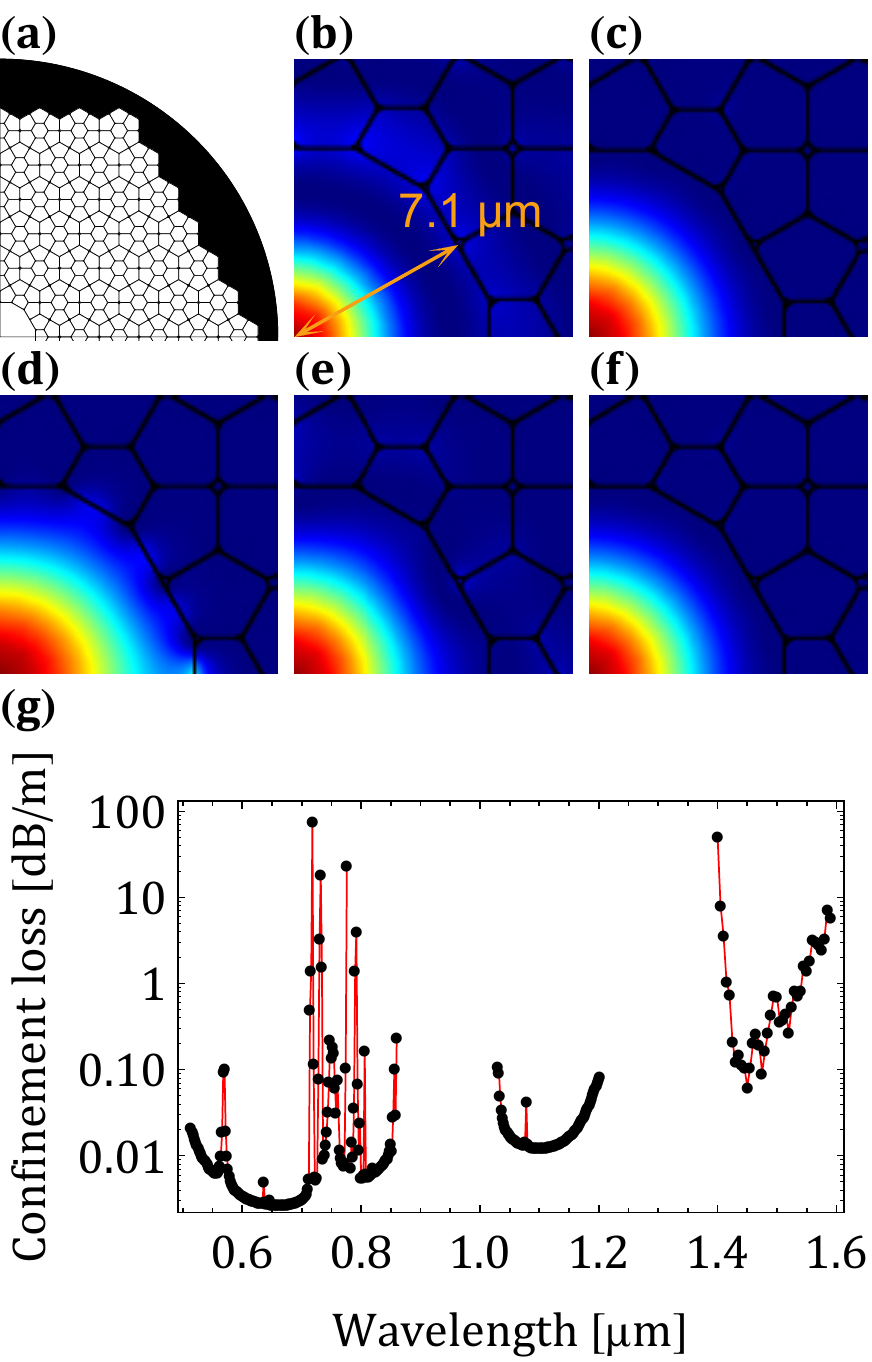}
\captionsetup{justification=justified}
\caption{(a)~[3\textsuperscript{6};3\textsuperscript{2}.4.3.4] HC-PCF layout. z component of the poynting vector of the (b) FM, (c) SH and (d) TH with the~[3\textsuperscript{6};3\textsuperscript{2}.4.3.4] structure. Mode spatial distributions at (e) ${1064~nm}$ and (f) ${800~nm}$ are also shown. (g) Confinement loss of the~[3\textsuperscript{6};3\textsuperscript{2}.4.3.4] HC-PCF.}
\label{fig:5}
\end{figure}

The effects of silica dispersion and the finite cladding were investigated with the finite element method (COMSOL Inc). Material dispersion was included with a three-term Sellmeier equation, absorption losses were included by adding to the silica regions a constant~${10^{-7}}$ imaginary refractive index~\cite{kitamura2007optical}. ${\Lambda=3~\mu m}$ was selected to design the fiber for a~${1500~nm}$ fundamental laser source (such as an Er/Yb fiber laser source). Strut thicknesses and apex curvatures were~${t=30~nm}$ and~${r=450~nm}$, respectively. The 19-cell core thickness and curvatures between the dodecagon core edges were identical to the structures strut thickness and apex curvatures. The~[3\textsuperscript{6};3\textsuperscript{2}.4.3.4] HC-PCF layout is shown in~Fig.~\hyperref[fig:5]{\ref*{fig:5}(a)}. With these parameters, guiding the FM, SH and TH is possible at~${\sim1570~nm}$, ${\sim785~nm}$ and~${\sim523~nm}$. Figs.~\hyperref[fig:5]{\ref*{fig:5}(b)}-\hyperref[fig:5]{\ref*{fig:5}(d)} show the z component of the poynting vector near the hollow-core at these wavelengths. The spatial overlap between the FM, SH and TH is exceptionally good. Intensity full width at half maximum~(FWHM) of the FM, SH and TH is~${\sim5.3~\mu m}$, ${\sim6.7~\mu m}$ and~${\sim8.4~\mu m}$, respectively. Confinement loss of the~[3\textsuperscript{6};3\textsuperscript{2}.4.3.4] HC-PCF is shown in~Fig.~\hyperref[fig:5]{\ref*{fig:5}(g)}. The confinement loss at the FM, SH and TH is~${\sim2.8~dB/m}$, ${\sim0.015~dB/m}$ and~${\sim0.015~dB/m}$, respectively.

The~${1064~nm}$ (Nd:YAG) and~${800~nm}$ (Ti:Sapphire) laser wavelengths are also supported with the~[3\textsuperscript{6};3\textsuperscript{2}.4.3.4] structure. Figs.~\hyperref[fig:5]{\ref*{fig:5}(e)} and~\hyperref[fig:5]{\ref*{fig:5}(f)} show the Gaussian mode distributions at these wavelengths. The confinement loss at~${1064~nm}$ and~${800~nm}$ is~${\sim0.014~dB/m}$ and~${\sim0.0056~dB/m}$, respectively. By scaling~${\Lambda}$, small blue and red shifts of all bandgaps are feasible. The struts thickness, which is the~[3\textsuperscript{6};3\textsuperscript{2}.4.3.4] structure smallest feature size, limits the blue shift of all bandgaps.

In conclusion, we demonstrated how to apply tiling and pattern theory in design of HC-PCFs for SH and TH guidance. Since the apexes of the hexagonal, square and triangular structures are located at the center of their tilings, it is possible to obtain a single cladding structure with multiple bandgaps by placing capillaries on the vertices of uniform or n-uniform tilings. Since the bandgap properties of high air-filling structures are mostly caused by the apexes, combining different apexes in a single structure can lead to claddings that support the FM, SH and TH. Utilizing the higher order bandgap of the triangular structure adds another restriction to the tilings, the desired structure should have struts of length~${\Lambda/\sqrt{3}}$. Many n-uniform tilings have been reported~\cite{galebach2018n}, this implies that there may be many more tilings that will allow to guide light at multiple wavelengths with HC-PCFs. For example, the 3-uniform~[3\textsuperscript{6};3\textsuperscript{3}4\textsuperscript{2};3\textsuperscript{2}4.3.4] and~[3\textsuperscript{6};3\textsuperscript{2}4.3.4;3\textsuperscript{2}4.3.4] tilings have square and triangular apexes and also have struts of length~${\Lambda/\sqrt{3}}$ (with p6m symmetry). As n increases, the n-uniform tilings have a larger unit cell size, thus computing the band diagrams will be much more challenging. For example, the rhombus unit cell of the~[3\textsuperscript{6};3\textsuperscript{2}.4.3.4] 2-uniform tiling (Fig.~\hyperref[fig:4]{\ref*{fig:4}(b)}) has an edge length of~${\Lambda(1+\sqrt{3})}$; yet, the~[3\textsuperscript{6};3\textsuperscript{3}4\textsuperscript{2};3\textsuperscript{2}4.3.4] and~[3\textsuperscript{6};3\textsuperscript{2}4.3.4;3\textsuperscript{2}4.3.4] 3-uniform tilings have a larger edge length of~${\Lambda(2+\sqrt{3})}$. Not only does the~[3\textsuperscript{6};3\textsuperscript{2}.4.3.4] structure support the SH and TH with a~${\sim1570~nm}$ fundamental laser source, it can also guide the wavelengths of Ti:Sapphire and Nd:YAG laser sources with low confinement losses.




\bibliographystyle{osajnlnt}
\bibliography{refs}





\end{document}